\documentclass[aps,prl,twocolumn,groupedaddress,longbibliography]{revtex4-1}
\usepackage{graphicx,amsmath,amssymb,color,xcolor}
\begin{document}

\title{Glassy polymers' strain-hardening moduli scale with their statistical-segment volumes}
\author{Robert S. Hoy}
\email{rshoy@usf.edu}
\affiliation{Department of Physics, University of South Florida}
\date{\today}
\begin{abstract}
Using molecular dynamics simulations, we show that a widely-accepted theoretical prediction for glassy-polymeric strain hardening moduli ($G_R \propto \rho_e$, where $\rho_e$ is the entanglement density) fails badly for semiflexible polymers with $N_e \lesssim 4C_\infty$.
By postulating that the length, energy and strain scales controlling $G_R$ are the Kuhn length $\ell_K$ and  statistical segment length $b = \sqrt{\ell_0 \ell_K}$ (where $\ell_0$ is the backbone bond length),  the intermonomer binding energy $u_0$, and the incremental elastic strain $\mathcal{S}_{\rm c}$ required to activate Kuhn-segment-scale plastic rearrangements, we develop a scaling theory predicting that $G_R = \mathcal{S}_{\rm c}(u_0/\ell_0^3) b^3$ in the athermal limit.
This prediction agrees quantitatively (semi-quantitatively) with simulated $G_R$ values for both flexible \textit{and} semiflexible polymer glasses subjected to athermal uniaxial-stress extension (constant-volume simple shear), over a range of $\ell_K/\ell_0$ that is wider than that spanned by real systems.
\end{abstract}
\maketitle

Understanding glassy-polymeric strain hardening has been a long-standing challenge.
Early efforts \cite{argon73a} focused on the observation that the normal-stress difference at strains well beyond yield takes the form $\Delta\sigma(\bar{\lambda}) \simeq  \sigma_{flow} +  G_R g(\bar{\lambda})$, where $\sigma_{\rm flow}$ is the plastic flow stress, $G_R$ is the strain hardening modulus, and $\bar{\lambda}$ is the macroscopic stretch tensor.
Rubber-elasticity-based theories \cite{haward68,arruda93,arruda93b} assume that $g(\bar{\lambda}) = \partial\tilde{g}/\partial\bar{\lambda}$, where $\tilde{g}(\bar{\lambda})$ is a dimensionless function that describes the reduction in the system's configurational entropy.
These theories predict $G_R = \rho_e k_B T$ where $\rho_e$ is entanglement density, but they have many flaws which have been discussed at great length in the literature.
In particular, they dramatically fail to predict both the magnitude and the temperature dependence of glassy-polymeric $G_R$  \cite{haward87,haward93,vanMelick03,kramer05,haward97}.

Noting that fitting stress-strain curves for various polymer glasses to $\Delta\sigma(\bar{\lambda}) \simeq  \sigma_{flow} +  \rho_e k_B T g(\bar{\lambda})$ produces values of $\rho_e$ that are up to two orders of magnitude larger than estimates from the melt plateau modulus $G_N^0 = 4\rho k_B T/5N_e$, Haward pointed out that the same functional forms $g(\bar\lambda)$ dominate the large-strain response of hyperelastic (neo-Hookean) materials \cite{haward87}.
He later noted \cite{haward93} that the prefactors $C$ obtained when fitting $G_R = C\rho_e k_B T$ to experimental data for $G_R$ are strongly chemistry-dependent, and suggested that $G_R$ is primarily determined by polymer glasses' local Kuhn-segment-scale structure rather than by entanglements.
Since then, a great deal of experimental and theoretical work has supported the idea that strain hardening is closely related to plastic flow and hence to local inter-segmental interactions; see \cite{argon13,rothbook16} for recent reviews.

The idea that strain hardening is only secondarily related to entanglement, however, remains controversial.
van Melick \textit{et al.}'s careful study of PS-PPO blends and variably-crosslinked PS showed that systems' measured $G_R$ were  linearly proportional to the $\rho_e$ obtained from independent measurements of their parent melts' plateau moduli $G_N^0$ \cite{vanMelick03}.
Simulations also found that $G_R \propto \rho_e$ \cite{hoy06,hoy07,hoy08}, and since there has (to date) been little to no experimental evidence to contradict this notion, it remains widely accepted.
As a consequence, because a \textit{simple} theory that quantitatively relates $G_R$ to microscopic structural parameters has remained elusive, it is tempting to assume that $G_R$ depends on these parameters in the same way that $G_N^0$ does.
For example, since the packing model  \cite{lin87,kavassallis87} asserts that $G_N^0\ell_K^3/k_B T \sim (\ell_K/p)^3$ [where $p = (\rho \ell_0 \ell_K)^{-1}$ is the packing length] and correctly predicts the $G_N^0$ values for all flexible polymers \cite{fetters94}, one could assume that these polymers' $G_R \sim G_N^0 \sim p^{-3} \sim (\rho \ell_0 \ell_K)^3$.
This hypothesis, however, has not been systematically investigated.
Experimental tests are impeded by the chemistry-dependent $C$, and previous comprehensive simulations focusing on strain hardening (e.g. Refs.\ \cite{hoy06,hoy07,hoy08})  have examined only a narrow range of $\ell_K$.

Moreover, recent studies have shown that $G_N^0 \sim p^{-3}$ scaling breaks down as $\ell_K/p$ increases into the semiflexible regime where $N_e \simeq C_\infty$ \cite{fenton22,milner20,hoy20,dietz22b}.
The semiflexible conjugated polymers (SCPs) that populate this regime are currently attracting great interest owing to their unique combination of electronic and mechanical properties \cite{rivnay13,liao15,xie18,xie20,fenton22,paleti24}. 
Recent molecular dynamics simulations have suggested \cite{dietz22c,nan23b} that these polymers can form ductile glasses, contrary to previous expectations \cite{kramer83}.
However, their strain-hardening behavior has not yet been systematically investigated, raising the question of whether they obey the typical $G_R \propto G_N^0 \propto \rho_e$ trend.

In this Letter, by re-analyzing data from Ref.\ \cite{nan23b}, we show that they do not.
The linear $G_R \sim \rho_e$ scaling breaks down as $\ell_K/p$ increases into the semiflexible regime; in this regime, $G_R$ is supra-linear in $\rho_e$.
On the other hand, we find that (at least for uniaxial-stress extension) $G_R \propto (\ell_K/\ell_0)^{3/2}$ to within our statistical uncertainties on $G_R$ and $\ell_K$, over a range of $\ell_K/\ell_0$  values that is wider than the range spanned by real systems \cite{fenton22,mark07}.
By assuming (i) that polymer glasses act like neo-Hookean solids \cite{haward87,haward93}, and (ii) that the range of their elastic intermonomer interactions is $b$, or equivalently that their plastic rearrangements mobilize a volume of order $b^3$, we formulate a simple scaling theory that predicts both the $3/2$ power law and the prefactor $G_0$ in the observed $G_R = G_0 (\ell_K/\ell_0)^{3/2}$ relationship to within a factor of order one.
In particular, we argue that
\begin{equation}
\Delta \sigma = \sigma_{\rm flow} +  \mathcal{S}_{\rm c}(u_0/\ell_0^3) (\ell_K/\ell_0)^{3/2} g(\bar\lambda) 
\label{eq:newlaw}
\end{equation}
in the athermal limit, where $\mathcal{S}_{\rm c}$ is the incremental \textit{elastic} strain required to activate these rearrangements.

All simulations were performed using LAMMPS \cite{thompson22}, and employed the semiflexible variant of the Kremer-Grest model \cite{kremer90,faller99}.
Monomers have mass $m$ and interact via the truncated and shifted Lennard-Jones potential $U_\textrm{LJ}(r) = 4u_0[(a/r)^{12} - (a/r)^{6} - (a/r_{c})^{12} + (a/r_c)^{6}]$, where $u_0$ is the intermonomer binding energy, $a$ is the monomer diameter, and $r_c = 2^{7/6}a$ is the cutoff radius. 
The covalent bond length $\ell_0 \simeq 0.96a$.
Variable chain stiffness is modeled using the standard potential $U_{\rm ang}(\theta) = \kappa u_0[1 - cos(\theta)]$, where $\theta$ is the angle between consecutive covalent-bond vectors and is zero for straight trimers.
All systems were equilibrated under standard  melt conditions (monomer number density $\rho = 0.85/a^3$, temperature $T = 1.0u_0/k_B$ \cite{kremer90})
as described in Ref.\ \cite{dietz22}.
Then they were slowly cooled to $T = 0$ (at zero pressure) as described in Ref.\ \cite{nguyen18}.
After cooling, systems were subjected to uniaxial-stress extension at a true strain rate $\dot\epsilon = 10^{-5}/\tau_{\rm LJ}$ or constant-volume simple shear at an engineering strain rate $\dot{\gamma} =  10^{-5}/\tau_{\rm LJ}$, where $\tau_{\rm LJ} = \sqrt{ma^2/u_0}$ is the Lennard-Jones time unit \cite{nan23b}.
Here we are focusing on the athermal ($T \to 0$) limit to isolate the dependence of $G_R$ on chain stiffness from its dependence on temperature.
In general, $G_R(\kappa,T) \simeq G_R(\kappa,0)[1 - T/T_g(\kappa)]$, where $T_g$ is the glass transition temperature \cite{vanMelick03,hoy06}.

Under standard  melt conditions, these polymers' equilibrium Kuhn lengths are approximately \cite{dietz22}
\begin{equation}
\begin{array}{c}
 \displaystyle\frac{\ell_K}{\ell_0} \approx  \displaystyle\frac{2 \kappa + \exp(-2 \kappa)  -1 }{1 - \exp(-2 \kappa) (2 \kappa + 1)} \\
 \\
+\  0.364 \times \left( \tanh\left[0.241 \kappa^2 - 1.73\kappa  + 2.08\right]+1\right) ,
\end{array}
\label{eq:ourkuhn}
\end{equation}
and their equilibrium entanglement lengths are \cite{dietz22}
\begin{equation}
N_e(\kappa) \simeq 80.5 - 70.4\tanh\left( \kappa/1.579 \right).
\label{eq:neinfofkap}
\end{equation}
The simulated  systems spanned nearly the entire range of positive $\kappa$ ($0.5 \leq \kappa \leq 5.5$) for which melts remain isotropic under these conditions \cite{faller99}, and a range of $N_e/C_\infty$ values ($1 \leq N_e/C_\infty \lesssim 28$) that is wider than the range spanned by real flexible and semiflexible polymer melts \cite{mark07,fenton22}.
Following standard procedure \cite{kramer83,hoy06}, we assume that the glasses' $N_e$ values are inherited from their parent melts and are given by Eq.\ \ref{eq:neinfofkap}.
Thus their entanglement \textit{densities} are $\rho_e(\kappa) = \rho_g(\kappa)/N_e(\kappa)$, where $\rho_g(\kappa)$ are their monomer number densities.

\begin{figure}[h]
\includegraphics[width=3.375in]{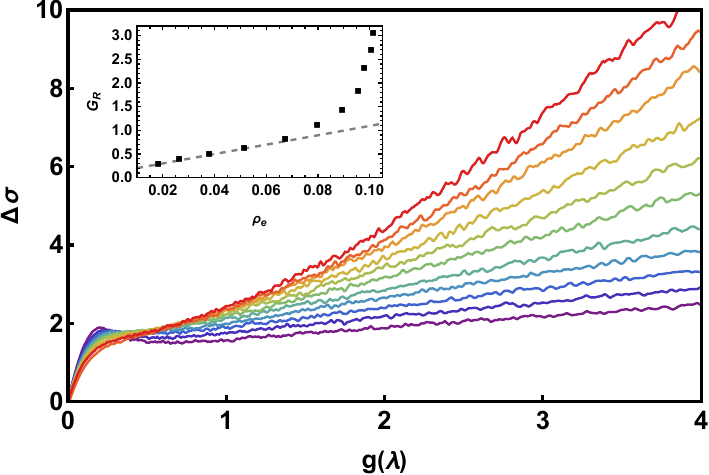}
\caption{Stress-strain curves for uniaxial-stress extension of Kremer-Grest glasses with $0.5 \leq \kappa \leq 5.5$.  Colors vary from purple to red with increasing $\kappa$.  The inset shows the $G_R(\kappa, 0)$ obtained by fitting these curves to Eq.\ \ref{eq:linGRfit} and a fit of the results for $\kappa \leq 2.0$ to the expected linear scaling with $\rho_e$ \cite{vanMelick03}.}
\label{fig:1}
\end{figure}

Uniaxial stress-strain curves for all systems are shown in Figure \ref{fig:1}.
Lower-$\kappa$ systems have larger initial elastic moduli and larger yield stresses owing to their more-efficient segmental packing \cite{nguyen18}, but all systems have comparable stresses in the plastic-flow regime.
As deformation continues, all systems exhibit a gradual crossover to standard  ``Gaussian'' strain hardening described by \cite{argon73a}
\begin{equation}
\Delta \sigma(\lambda) = \sigma_{\rm flow} + G_R g(\lambda);
\label{eq:linGRfit}
\end{equation}
here $\lambda = \exp(\epsilon)$ is the stretch along the tensile direction and  $g(\lambda) = \lambda^2 - 1/\lambda$ \footnote{This definition of $g(\lambda)$ ignores the very small increases in volume that occur at small strains, but correcting for these does not alter the main results presented below.}.
At even larger strains (not shown here), systems exhibit a crossover to nonlinear (Langevin \cite{haward68,arruda93b}) strain hardening that becomes more dramatic with increasing $\kappa$.
% \cite{hoy08}.
All of these stress-strain curves are qualitatively consistent with those reported in previous comprehensive simulations of glassy-polymeric strain hardening \cite{hoy06,hoy07,hoy08}; the only novel aspect is that here they span a much wider range of $\kappa$ and $\ell_K/\ell_0$.

We estimate all systems' $G_R(\kappa, 0)$ by fitting their stress-strain curves to Eq.\ \ref{eq:linGRfit} over the range  $2 \leq g(\lambda) \leq 4$.
The inset shows how the resulting $G_R$ values vary with $\rho_e(\kappa)$.
As expected \cite{vanMelick03,hoy06}, we find that $G_R$ is linearly proportional to $\rho_e$ for flexible polymer glasses.
More specifically, we find that $G_R \simeq c_0 + c_1\rho_e$ for systems with $0.5 \leq \kappa \leq 2.0$, and that while $c_0$ and $c_1$ are both sensitive to the range of $g(\lambda)$ used to estimate $G_R$, the overall result is robust.
For larger $\kappa$, however, the expected $G_R \sim \rho_e$ scaling breaks down.  
$G_R$ continues to increase steadily with increasing $\kappa$, whereas $d\rho_e/d\kappa$ \textit{decreases} steadily as $N_e(\kappa)$ approaches its minimum possible value ($N_e^{\rm min} \simeq C_\infty$) for systems lacking long-range nematic order \cite{hoy20,dietz22b}.
As a consequence, $dG_R/d\rho_e$ increases steadily and appears to diverge as $\kappa$ approaches its isotropic-nematic transition value $\kappa_{\rm in}$ (which is slightly above $5.5$ for the standard Kremer-Grest melt \cite{faller99}).
Notably, the range of $\kappa$ for which the linear scaling fails (i.e., $2.5 \leq \kappa \leq 5.5$) corresponds to the range $1 \leq N_e/C_\infty \lesssim 4$, i.e.\ the range of $N_e/C_\infty$ values corresponding to SCPs \cite{fenton22,dietz22b}.

The observation that $G_R$ is not always directly proportional to $\rho_e$ is an additional piece of evidence that entanglements play an indirect role in controlling strain hardening.
One obvious candidate for a structural feature that \textit{does}, in fact, directly control it is Kuhn segments \cite{haward93,vanMelick03,kramer05,haward97,argon13,rothbook16}.
Flory's definition of the characteristic ratio is 
\begin{equation}
C_\infty =  \left( \displaystyle\frac{1 + \langle \cos(\theta)\rangle}{1 - \langle \cos(\theta)\rangle} \right) \left( \displaystyle\frac{1 + \langle \cos(\psi)\rangle}{1 - \langle \cos(\psi)\rangle} \right), 
\label{eq:genKuhn}
\end{equation}
where $\psi$ is the dihedral angle formed by three consecutive covalent-bond vectors.
$C_\infty = \ell_K/\ell_0$ in \textit{undeformed} glasses, but it can increase dramatically over the course of applied deformation as $\langle \cos(\theta)\rangle$ and $\langle \cos(\psi)\rangle$ evolve.
It has long been known that tensile plastic deformation in polymer glasses occurs largely through irreversible dihedral transitions that tend to increase $\langle \cos(\psi)\rangle$ \cite{haward97}, and Ref.\ \cite{nan23b} showed that the ratio $C_\infty(\lambda)/C_\infty(1)$ closely tracks the deviatoric stress $\sigma_{dev}(\lambda)$ in the same bead-spring polymer glasses considered here.

Predicting how polymer glasses'  $\langle \cos(\theta)\rangle$ and $\langle \cos(\psi)\rangle$ evolve under applied deformation is very challenging, however, and therefore so is developing a Kuhn-segment-based theory of strain hardening using this approach.
The phenomenological theory recently developed by Merlette \textit{et al.}\ \cite{merlette23,merlette25} employs a closely related approach, attributing strain hardening  to the increasing free energy barriers for $\alpha$ relaxation that arise from the increasing local \textit{alignment} of Kuhn segments.
However, this theory does not explicitly address how $G_R$ varies with $\ell_K/\ell_0$, and indeed it predicts that the length scale controlling these free energy barriers ($\xi = 3-5\rm{nm}$) is considerably larger than $\ell_K$.
While it does predict that $G_R$ scales approximately linearly with the ``monomer'' volume $a^3$, it treats $a$ as an adjustable parameter rather than as a quantity that is precisely specified by polymer chemistry.

Another candidate  for the controlling structure is statistical segments, which (unlike Kuhn segments) are effectively rigid and hence lack dynamic internal variables like $\theta$ and $\psi$.
Statistical segments are the ``atoms'' of Chen and Schweizer's microscopic-physics-based polymer nonlinear Langevin equation (PNLE) theory of glassy polymers' thermal- and deformation-history-dependent mechanics \cite{chen08b,chen09,chen10b}. 
In this theory, the elastic shear modulus $G$ of a polymer glass is given by \cite{chen10b}
\begin{equation}
G = \displaystyle\frac{k_B T}{60\pi^2} \displaystyle\int_0^\infty \left[\displaystyle\frac{q^2 }{S(q)} \right]^2 \left( \displaystyle\frac{dS}{dq} \right)^2 \exp\left( -\displaystyle\frac{-q^2 r^2_{\rm loc}}{3S(q)} \right) dq ,
\label{eq:KK1}
\end{equation}
where $S(q)$ and $r_{\rm loc}$ are the segments' (thermal- and deformation-history-dependent) static structure factor and localization length.
PNLE predicts that strain hardening is a consequence of increasingly-anistropic single-chain and intermolecular correlations that suppress long-wavelength density fluctuations and hence increase $G$.
It predicts $G \sim k_BT/b^3$ for \textit{fixed} $S(qb)$ and $r_{\rm loc}/b$, but since both of these measures are strongly deformation-dependent, it does not predict a simple universal relationship between $G_R$ and $b$.

Here we take an alternative approach that is much less rigorous than those employed in Refs.\  \cite{merlette23,merlette25,chen08b,chen09,chen10b} but allows us to make such a prediction.
Consider a semiflexible pearl-necklace-like polymer where the monomer diameter and covalent bond length are both $\ell_0$ and the intermonomer binding energy is $u_0$.
Since statistical segments are effectively rigid, it is reasonable to assume that the range of elastic intermonomer interactions is $b$, and therefore that the total elastic strain energy \textit{per monomer} in a deformed glass composed of these polymers scales as $u_0(b/\ell_0)^3$.
The associated  energy density is $\mathcal{E} \sim u_0b^3/\ell_0^6$.
Then, assuming that these glasses act like neo-Hookean solids and their strain energy density is $U \sim \mathcal{E} \tilde{g}(\bar\lambda)$ leads to a normal stress difference $\Delta\sigma \sim \mathcal{E} g(\bar\lambda)$ consistent with Gaussian strain hardening, with $G_R \sim u_0 b^3/\ell_0^6 \sim (u_0/\ell_0^3) (\ell_K/\ell_0)^{3/2}$.

\begin{figure}[h]
\includegraphics[width=3.375in]{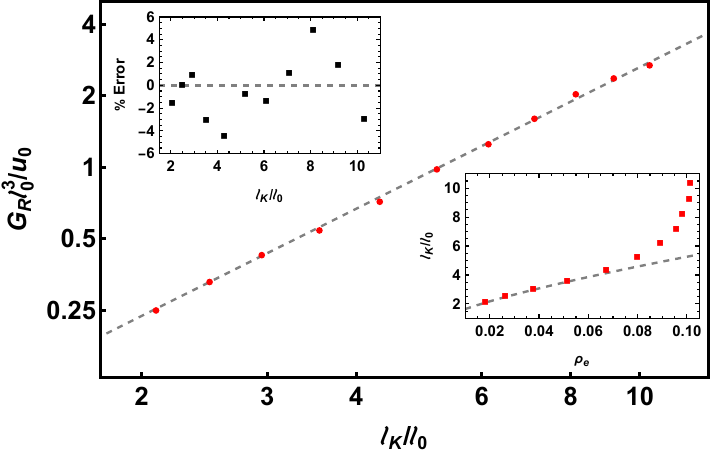}
\caption{Data for the scaled strain hardening moduli $G_R(\ell_K)/(u_0/\ell_0^3)$ (red points) and a fit to $G_R =  S_{\rm C}(u_0/ell_0)^3(\ell_K/\ell_0)^{3/2}$ with $S_{\rm C} = 0.08364$ (gray line).  The upper and lower insets respectively show the fractional errors associated with this fit and a plot of $\ell_K/\ell_0$ vs.\ $\rho_e$.}
\label{fig:2}
\end{figure}

Obviously glassy polymers are \textit{not} elastic solids; the majority of the work done in deforming them beyond their yield points is dissipated as heat \cite{hasan93,senden12c}.
A simple interpretation of the prediction $G_R \propto b^3$ that is consistent with this experimental fact is that $G_R$ is proportional to the number of non-covalent inter-monomer bonds that must be broken for a Kuhn volume to plastically re-arrange via irreversible changes in its constituent chains' $\{ \theta \}$ and $\{ \psi \}$.
This interpretation is consistent with simulations that have shown that the rate of non-covalent bond breaking is approximately linear in $g(\lambda)$ and [for fixed $g(\lambda)$] increases with increasing $\ell_K$ \cite{hoy07,hoy08}.
It specifies the scaling, but not the magnitude, of $G_R$.

Inspired by Frenkel's classic yield criterion ($\tau_{\rm Y} = \mu/5$, where $\tau_{\rm Y}$ is the ideal strength of a material and $\mu$ is its shear modulus \cite{frenkel26b}) as well as more recent work showing that metallic glasses' $\tau_{\rm Y} \simeq \gamma_{\rm c}\mu$ (where $\gamma_{\rm c} \simeq 1/37.5$ is a critical shear strain \cite{johnson05}), we postulate that $G_0 = \mathcal{S}_{\rm c}(u_0/\ell_0^3)$ with $\gamma_{\rm c} < \mathcal{S}_{\rm c} \leq 1/5$.
Here $\mathcal{S}_{\rm c}$ is the incremental \textit{elastic} strain required to activate the abovementioned cooperative 
%statistical- and 
Kuhn-segment-scale plastic rearragements, and $b^3$ is the volume of the associated shear transformation zones \cite{argon79}.
$\mathcal{S}_{\rm c}$ should be larger than $\gamma_{\rm c}$ because polymer glasses' yield strains are substantially larger than those of metallic glasses \cite{haward97}.
In this picture, Eq.\ \ref{eq:newlaw} captures how the normal stress difference required to activate these rearrangements increases with large-scale chain orientation.
This could offer a particularly simple explanation for why $G_R$ values can be $\sim 10^2$ times larger than $\rho_e k_BT$ even for $T$ that are relatively near to $T_g$ \cite{vanMelick03,kramer05}.
Specifically, $G_R \gg \rho_e k_BT$ in these systems because the free energy density that sets the magnitude of $G_R$ is $\mathcal{S}_{\rm C}(u_0/\ell_0^3)(b/\ell_0)^3$ rather than $\rho_e k_B T$.

Figure \ref{fig:2} plots the scaled moduli $G_R(\kappa,0)\ell_0^3/u_0$ obtained from our molecular dynamics simulations versus $\ell_K/\ell_0$.
Remarkably, all data fall on the line $G_R \ell_0^3/u_0 \simeq 0.084(\ell_K/\ell_0)^{3/2}$ to within our estimated statistical and systematic errors.
The fractional deviations between the fit and measured values of $G_R$ are all less than 6\%, which is comparable to the deviations from the predicted $\ell_K(\kappa)$ (Eq.\ \ref{eq:ourkuhn}) reported in Ref.\ \cite{dietz22} and the dependence of the ratios of different $G_R$ on the range of $g(\lambda)$ used to fit the stress-strain curves to Eq.\ \ref{eq:linGRfit}.
Moreover, the prefactor ($0.084$) lies well within our estimated range for $\mathcal{S}_{\rm c}$.

\begin{figure}[h]
\includegraphics[width=3in]{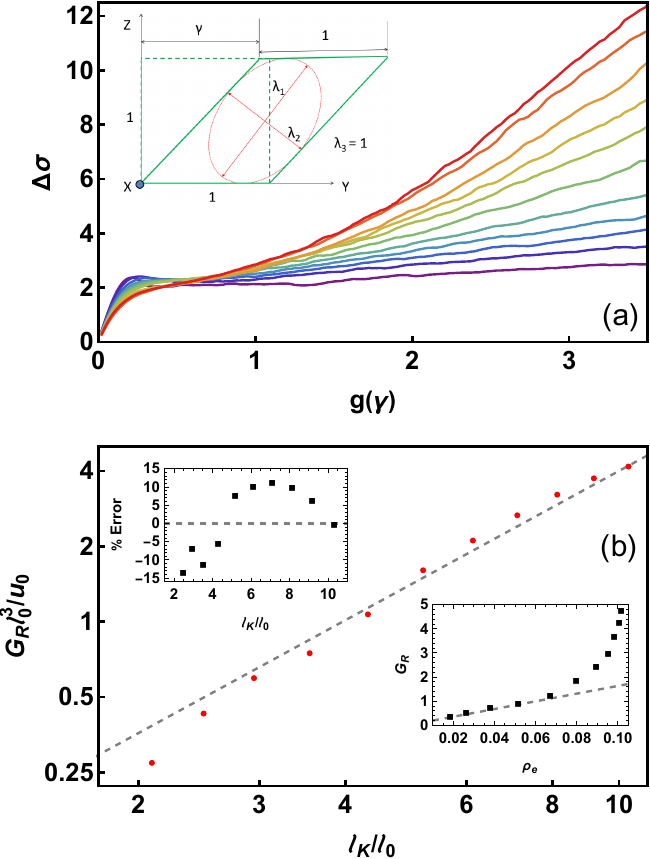}
\caption{Constant-volume simple-shear response for the systems analyzed in Figs.\ \ref{fig:1}-\ref{fig:2}.
Panel (a) shows the stress-strain curves; the inset schematically illustrates the principal stretch directions. 
Panel (b) shows the $G_R(\kappa, 0)$ obtained by fitting these curves to Eq.\ \ref{eq:GRshear} along with a fit to $G_R = G_0 (\ell_K/\ell_0)^{3/2}$ scaling.
The upper-left inset shows the fractional errors associated with this fit,  while the lower-right inset shows a fit of the results for $\kappa \leq 2.0$ to the expected $G_R \sim \rho_e$ scaling \cite{vanMelick03}.}
\label{fig:3}
\end{figure}

Are the theoretical arguments leading to Eq.\ \ref{eq:newlaw} robust?
Constant-volume simple shear provides a more stringent test because stresses are triaxial and include both tensile and compressive components.
As a consequence, systems exhibit a greater tendency to deform inhomogeneously (via shear-banding \cite{kramer83,haward97}).
For our deformation protocol \cite{nguyen18,nan23b}, the principal tensile and compressive stretches are respectively $\lambda_1=(\sqrt{\gamma^2+4} + \gamma)/2$ and
$\lambda_2=(\sqrt{\gamma^2+4}-\gamma)/2$, with associated principal stresses  
\begin{equation}
\sigma_{1,2}= (\sigma_{yy}+\sigma_{zz})/2 \pm \sqrt{(\sigma_{yy}-\sigma_{zz})^2/4+\sigma_{yz}^2}.
\label{eq:princstress}
\end{equation}
The normal stress difference is $\Delta\sigma \equiv \sigma_1 - \sigma_2$, and Gaussian strain hardening corresponds to
\begin{equation}
\Delta\sigma(\gamma) = \sigma_{\rm flow} + G_R g(\gamma),
\label{eq:GRshear}
\end{equation}
where $g(\gamma) = \lambda_{1}^2 - \lambda_2^2 \equiv \gamma\sqrt{4 + \gamma^2}$.
Stress-strain curves for all systems are shown in  Figure \ref{fig:3}(a).
$\kappa = 0.5$ systems shear-band upon yielding as indicated by the long post-yield plateau in their $\Delta\sigma(g)$, but all other systems exhibit a gradual crossover to Gaussian hardening over the range $0.5 \lesssim g(\gamma) \lesssim 2$ that is similar to that shown in Fig.\ \ref{fig:1}.

If our theoretical picture is correct, we would expect to find  that $\Delta \sigma(\gamma)  = \sigma_{\rm flow} +  \mathcal{S}_{\rm c}(u_0/\ell_0^3) (\ell_K/\ell_0)^{3/2} g(\gamma)$, where $\mathcal{S}_{\rm c}$ can differ from the value obtained above (owing to the different strain geometry), but only by a factor of order one.
Fig.\ \ref{fig:3}(b) shows the $G_R$ obtained by fitting these stress-strain curves to Eq.\ \ref{eq:GRshear} over the range $2 \leq g(\gamma) \leq 3.5$.
Once again, the classic $G_R \sim \rho_e$ scaling holds for flexible polymers but breaks down for semiflexible polymers.
For $\kappa > 0.5$, the hardening moduli instead are (once again) roughly linear in $b^3$.

Fitting these moduli to $G_R = S_{\rm C}(u_0/\ell_0^3)(\ell_K/\ell_0)^{3/2}$ yields $S_{\rm C} \simeq 0.127$, a value that is both within our estimated range for $\mathcal{S}_{\rm c}$ and within an $\mathcal{O}(1)$ factor of the value obtained for uniaxial-stress tension.
However, in contrast to the unaxial-stress case, the agreement is only \textit{semi}-quantitative.
The rms fractional deviation between the fit and measured $G_R$ for $\kappa > 0.5$ is 12.6\%, which is five times larger than the corresponding deviation for uniaxial stress.
This deviation is too large to be explained by the abovementioned statistical and systematic uncertainties in $\ell_K$ and $G_R$.
We expect that it arises from other factors  that are not captured by our simple model, e.g.\ the greater tendency for heterogeneous deformation and earlier onset of Langevin hardening in simple shear \cite{arruda93b}.
Nevertheless, $G_R \sim (\ell_K/\ell_0)^{3/2}$ scaling clearly describes this dataset far better than $G_R \sim \rho_e$ scaling does.

The lower-right inset in Fig.\ \ref{fig:2} shows that $\ell_K/\ell_0$ (much like $G_R$) increases linearly with $\rho_e$ for  $\kappa \leq 2.0$ but faster than linearly for $\kappa \gtrsim 2.5$.
This suggests that the failure of $G_R \sim \rho_e$ scaling can be physically interpreted as follows: $G_R$ is actually scaling with another quantity (specifically, $\ell_K/\ell_0$) that itself scales linearly with $\rho_e$ \textit{for flexible systems}, and therefore the $G_R \sim \rho_e$ scaling observed \textit{for these systems} indicates correlation rather than cause.
van Melick \textit{et al.}'s results for PS/PPO blends \cite{vanMelick03} are consistent with this interpretation because the PPO weight fraction $f_{\rm PPO}$ (which itself scales linearly with $\rho_e$) controls these systems' average segment-scale structure and interactions, and hence their effective $\langle (u_0/\ell_0)^3 b^3 \rangle$.
Their demonstration that  $G_R \propto (\rho_e + \rho_x)$ in pure PS with a wide range of crosslink densities $\rho_x$ is harder to rationalize since these systems had nearly-fixed $b$ and $u_0$, but it could also be consistent with Eq.\ \ref{eq:newlaw} if crosslinks increase $\mathcal{S}_{\rm c}$.
Such increases would be consistent with (or, at least, do not conflict with) Ref.\  \cite{vanMelick03}'s hypothesis that crosslinks increase $G_R$ by reducing segmental \textit{mobility}.

 This interpretation is also naturally consistent with the notion that entanglements play only an indirect role in controlling strain hardening, and specifically with the hypothesis that by forcing chains to deform affinely on chemical length scales $n > N_e$, they make polymer glasses neo-Hookean.
While this idea was proposed long ago \cite{haward87,haward93} and is assumed to be correct in modern theories of glassy polymer mechanics  \cite{chen08b,chen09,chen10b,merlette23,merlette25}, it had previously been difficult to demonstrate conclusively in the absence of evidence that $G_R \propto \rho_e$ scaling is non-universal in \textit{neat} polymer glasses.
Violations of this scaling were previously reported  in simulations of binary mixtures of long ($N \gg N_e$) and short ($N \ll N_e$) chains which found that $G_R$ was linear (while $\rho_e$ was quadratic) in the long chains' weight fraction $f$ \cite{hoy09b}, but the conclusions of that study have been criticized on the grounds that such mixtures are brittle in experiments unless $1-f \ll 1$ \cite{liu15c,liu19b}.

In conclusion, we have shown in this Letter that $G_R \propto \rho_e$ scaling breaks down as $N_e/C_\infty$ decreases into the regime corresponding to SCPs \cite{rivnay13,liao15,xie18,xie20,fenton22,paleti24}.
We have also taken a first step torwards resolving this issue by formulating a simple 
%theoretical expression (Eq.\ \ref{eq:newlaw})
theory that quantitatively predicts $G_R$ from systems' statistical- and Kuhn-segment-scale structure and interactions, over a range of $N_e/C_\infty$ ($1 \lesssim N_e/C_\infty \lesssim 28$) that is wider than spanned by real systems \cite{fenton22,mark07}.
The theory in its current form is limited by its athermal nature, but one could imagine extending it to account for thermalization of the Kuhn-segment-scale plastic rearrangements and hence to predict how $G_0$ depends on both temperature and strain rate \cite{johnson05}.

We thank Kang Chen, Kenneth Schweizer, and Didier Long for helpful discussions.
This material is based upon work supported by the National Science Foundation under Grant No. DMR-2026271.

%\bibliography{/Users/rshoy/Documents/allref} 

\end{document}